# Powers of the Universe: Empowering primary school students with the powers of ten notation

Anastasia Lonshakova [a], David G. Blair[a], David F.Treagust [b], Marjan Zadnik [a]

[a] Department of Physics, The University of Western Australia, Perth WA 6009, Australia

[b] STEM Education Research Group, School of Education, Curtin University, Perth WA 6102, Australia

00106381@uwa.edu.au

**Abstract**

Numbers, both very large and very small, are crucially important for understanding the modern world. This paper assesses trials of a mathematics and physics module called *Powers of the Universe* in which arithmetic with extreme numbers (large and small) is developed through early learning of the powers of ten notation. We trialled a 6-hour progression of lessons based on activities and group learning with students aged 7-13 years. We measured students' ability to estimate, compare and calculate extreme numbers using pre and post-tests to evaluate the program. Results demonstrated students' strong enthusiasm and positive learning outcomes in areas normally assumed to be beyond the capability of students in this age group. We discuss the age dependence of some results and suggest an optimum strategy for enhancing primary school mathematics. The module has been delivered, as part of a broader five-module program called Maths for Einstein's Universe, that aims to reduce maths anxiety through programs with direct relevance to the modern world and reduced emphasis on exactness.

# Introduction and Literature review

The need for making mathematics education more relevant to students through integration of knowledge and skills connected to the modern understanding of physical reality into highlighted by several authors (Maass, 2019; Welch, 2016). In response to this need, we have developed a program called *Maths for Einstein's Universe* that aims to introduce students at an early age to broad range of mathematical concepts from vectors to probability and curved space geometry at an early age (Authors, 2023). The idea is to reduce emphasis on arithmetic, and open students' minds to a broader range of mathematical concepts that are both relevant and connected to topics students find interesting. Among the essential skills for the modern world is the proficiency with extreme numbers (Resnick et al.,2017). This paper describes one component of *Maths for Einstein's Universe*, called *Powers of the Universe. It* aims to develop understanding of extreme numbers, large and small, at an early age.

Extreme numbers are a part of our modern world. Indeed, we are frequently exposed to extreme numbers outside human everyday experience relating to topics of important national discourse such as climate change science, the global population, stock exchange data, energy projects and computer power (Maass et al., 2019). Astronomical discoveries and discoveries of ancient fossils also involve enormous numbers to describe distances and periods of time. Many of these concepts capture students' interest. Similarly, small numbers have enormous importance over a diverse range of topics from nanotechnology to viruses and the size of transistors in computer chips. The sizes of micro-organisms, atoms, molecules, and sub-atomic particles are part of the modern science experience, down to tiny vibrations of space that are measured by gravitational wave detectors when distant black holes coalesce. In the time of the global pandemic, learning about both the small size of viruses and the enormous speed of exponential spreading are vitally important.

*Widespread miscomprehension of extreme numbers*. There is widespread recognition of the need for improved numerical literacy (Maass et al., 2019). For most people, their intuitive understanding of scale is very poorly developed (Landay et al., 2013). For many people, big numbers are just words with very little meaning (Resnick et al., 2017, Trueborn & Landy, 2010). This leads to many examples in the mass media where very few people notice enormous errors in magnitude when, for example, reporters confuse millions and billions. Difficulties in comprehension of big and small numbers have been recognized in published papers (Swart et al., 2011), as students represented their understanding of ''size and scale'', particularly in nanoscale. There are recognized difficulties in comparing magnitudes of extreme size (Nataraj et al., 2012) and in comprehending extreme time intervals (Libarkin et al., 2007). The reason for this is that as numbers increase in magnitude, they become progressively less distinguishable to humans. Furthermore, context-based learning was reported as a very effective method of enhancing number sense (Beswick, 2011).

**Extreme numbers in astronomy, geology, nanoscience and real life.** Previous research has investigated the abilities of students to compare the sizes and distances of astronomical objects (Rajpaul et al, 2018). There is other evidence, that the progressive alignment of numbers on the linear number line successfully produced improvement of estimations on increasingly larger scales (Thompson, 2010). The use of ranking exercises with unfamiliar numbers has been reported as an efficient method for developing an understanding of astronomical distances or sizes. In a ranking task, students are required to arrange a set of objects in a specific order based on certain criteria. For instance, in Rajpaul et al. (2018) students were tasked with ranking five objects - galaxy, planet, star, universe, and solar system - from smallest to largest. This study presents evidence of how the comprehension of large numbers can be addressed in a specific context involving astronomical objects. Understanding the scale of the universe from the smallest objects to the most distant objects is a significant challenge. Jones et al. (2007) investigated how the understanding of scale was improved after watching the Powers of Ten movie (1968) in which students watched constantly increasing/ decreasing distances in the Universe in powers of ten. After the visualisation, students improved their capability to put sizes on the number line using scientific notation.

The problem of understanding big numbers in geology has been addressed in several studies (Cheek, 2012; Resnick et al., 2017). In these research studies, the geological time scale was introduced to students through spatial distribution on the linear number line to comprehend the differences between time periods. For example, students can unroll a roll of bathroom tissue to depict the 4.6-billion-year history of Earth. As the teacher discusses the various events in Earth's history, students are encouraged to hold the tissue timeline at certain points (Clary, 2009). The research indicates that employing a spatial number distribution significantly enhances students' comprehension of extreme numbers. However, the linear arrangement does not allow detailed exploration of modern historical events.

Concepts of small size and small scale are important for nanotechnology. By investigating students understanding of the nanoscale through proportional reasoning and ranking tasks, Swart (2001) showed that linear and logarithmic proportional reasoning serves as a valuable stage toward a more thorough comprehension of extreme scales. However, it is important to note this study was conducted with undergraduate engineering students, who may have already had conceptions of extreme scales.

The study by Albarracín and Gorgorió (2019) explored students' understanding of large numbers through real-life estimation problems. For instance, students were asked to estimate the number of people that could fit in a given surface area such as a large open space on the first open floor of a school building (called a porch by the authors). Students approached this task by dividing the large area into smaller sections and making approximations for how many smaller sections fit the whole area. This method allowed students to grasp large numbers within a specific context, by employing an even distribution of elements within the given area.

Most of the reported approaches use a linear representation of numbers, which seems to be limited on extreme scales as huge intervals between numbers are needed.

*Extreme Number reasoning.* There is clear evidence that reasoning about large numbers and small numbers is different from reasoning about "ordinary" numbers that people experience in everyday life (Resnick, 2017). People have to stretch their understanding of ordinary magnitudes to the area of unfamiliar quantities. For example, they divide large numbers into categories (millions, billions etc) and stretch linear responses across items within each category (Landay et al., 2013), including relational reasoning (Resnick et al., 2017) or use spatial analogies (Clary, 2009). Despite numerous approaches to address and investigate the issue of understanding extreme numbers (Harkness & Brass, 2022), the difficulty problem persists.

*Linear and Logarithmic mental number lines.* Usually, people overestimate small magnitudes and underestimate big magnitudes (Resnick, 2017), showing that they do not comprehend them correctly. Understanding numbers is directly related to the accurate positioning of numbers on the mental number line (Booth & Siegler, 2006). The understanding of magnitudes outside human experience is indicated by the accurate placement of the numbers on the mental number line with the correct distance between them. Powers of ten and logarithmic thinking are the only way to effectively represent both small and big numbers on the number line (Fig 1).

**Fig.1**

*The log number line and the linear line* are *shown matching at the value 1.*

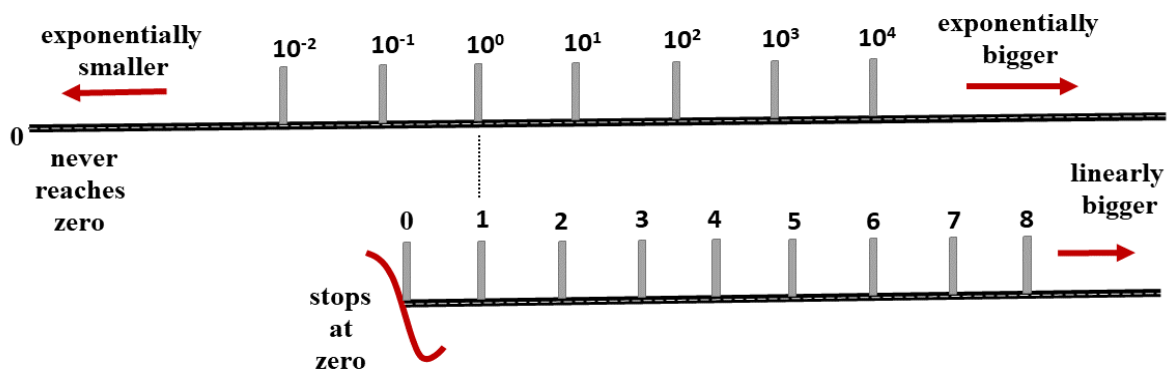

The logarithmic number line does not have a zero. As negative powers increase in magnitude the values exponentially approach the unattainable zero. The logarithmic number line allows us to represent the enormous range of scales we experience in the universe, which is impossible for the linear line.

If we are to represent the scales of physical reality, we need to facilitate the transition of the mental number line from linear to logarithmic (Moeller et al., 2009). However, logarithms are recognised to be problematic in the secondary school curriculum (Weber, 2002) and students' difficulties with operations involving exponents and logarithms in the middle school years have been widely investigated (Ulusoy, 2019). Even teachers struggle to comprehend exponential growth (Ellis et al., 2016).

Nevertheless, the research on the numerical cognition of young children shows that they have a logarithmic-type of thinking. For example, if asked to distribute numbers on the line from 0 to 100, students place 10 near the middle, devoting more space to the smaller numbers (Dehaene et al., 2008). This perceptual preference of logarithmic mapping of numbers on the number line may be connected to Weber-Frechner's law about logarithmic sensation (Dehaene, 2003). Initially, this law described perception of loudness and brightness etc. Later Stevens and Shepard (1975) proposed extending this law to include abstract thinking such as the sense of numbers This suggestion means that the inclination towards a non-linear spatial arrangement of numbers on the mental number line could be neurobiologically based. This means that we can utilize young children's natural tendency or perceptual preferences to place numbers on the number line to facilitate the development of logarithmic reasoning. We hypothesize that in this case, a logarithmic number line need not be difficult to acquire as a tool.

There is evidence that powers of ten, and hence the logarithmic scale, was known in antiquity. They were first used in India 3000 years ago. The Veda called Vajasaneyi Samhita lists the words used for various powers of ten, as listed in Table 1 (Nataraj & Thomas, 2012).

**Table 1.**

*Number Names for Powers of Ten*

| Eka | 1 | Sahasra | $10^3$ | Prayuta | $10^6$ | Madhya | $10^{10}$ |
|---|---|---|---|---|---|---|---|
| Dasa | 10 | Ayuta | $10^4$ | Arbuda | $10^7$ | Anta | $10^{11}$ |
| Sata | $10^2$ | Niyuta | $10^5$ | Nyarbuda | $10^8$ | Parardha | $10^{12}$… |
| | | | | Samudra | $10^9$ | Loka | $10^{19}$ |

Also, about 1000 years later, the Buddhist work called Lalitavistra quotes Gautama Buddha's replies to the mathematician's question about counting with extreme numbers. This extended names of individual powers of ten to $10^{53}$, a number called talaksana. At that time, we presume that such extreme numbers could only have been of abstract significance.

Recognition of these historical discoveries emphasises the idea that the base-ten logarithmic system is not a product of contemporary mathematics needed for understanding the scale of the universe, but a numerical system for representing scale that greatly pre-dates modern scientific discoveries. Presumably this was a product of prodigious imagination, including speculations about extreme durations of time found in the Hindu scriptures such as a maha-kalpa (life of Brahma). In this case, the concept of extreme numbers has been part of human civilisation for several millennia.

After investigating the numerical skills of students in different indigenous cultures, who demonstrated the logarithmical distribution of numbers on the scale, researchers have concluded that a linear number line may be a product of culture and formal education (Dehaene, 2008) associated with the use of rulers and graphs.

Siegler et al. (2009) present a contrary view, suggesting that logarithmic thinking is only a step towards "advanced" linear reasoning. They propose that logarithmic thinking is manifested in children's reasoning in the early stage of brain development, and that it progresses to linear thinking with age and experience. This argument would imply that the exclusive emphasis of linear numerical representations in primary school education makes the development of logarithmic thinking skills more challenging in later years.

Other researcher have investigated the changes in mental number representation that occur during the development of children (Opher et al., 2010).  A finding from a highly cited paper (Dehaene, 2008) has shown that the shift from logarithmic to linear mental number line occurs between first and fourth grade. This observation implies that it may be important to start learning powers of ten in primary schools, while students retain the natural ability to think logarithmically. Mahajan (2018) suggests that this approach could bring remarkable benefits for developing mathematical skills in general.

The above arguments suggest that early introduction of logarithmic reasoning that progresses through the whole curriculum may be important to give students the best ability to deal with the extreme numbers that are relevant in the modern world.
Whether or not logarithmic thinking is 'natural' in terms of brain development, the above arguments do not offer discouragement. Powers of ten is the best tool we have available to comprehend the enormous scale of the Universe as well as many aspects in the modern world (Bitterly et al.,2022).

## The goal of this research

This paper presents the design, development and trials of an activity-based module called *Powers of the Universe.* Our aim was to introduce students aged 7 to 13 years to powers of ten, assess their logarithmic reasoning abilities and determine the best age for introducing these concepts. To make it accessible to younger students required innovative approaches that would provide comprehension of extreme numbers while developing proficiency in using powers of ten as a mathematical tool. Adequate comprehension of extreme numbers should include:

1) An ability to position and read numbers on the logarithmic number line both mentally and physically.

2) A sense of proportionality between pairs of numbers on the logarithmic number line

3) Ability to perform mathematical operations with simple powers of ten.

4) An ability to make estimations by rounding numbers to appropriate powers of ten.

*To achieve the above, we relied* on the well-established effectiveness of context-based learning, as highlighted by Beswick (2011).

In the next section we consider the use of powers of ten as a mathematical tool for enhancing the understanding of extreme numbers. Then we will present the activity-based learning sequences used in this study, followed by the methodology for the module development, testing and analysis of results used for assessing its effectiveness. We will conclude by discussing the effectiveness of our trials and identifying limitations and and proposed future developments.

## Powers of ten as a notational tool

In 2007 a highly cited paper by Ashcroft and Krause (2007) showed that working memory is increasingly involved in problem-solving as the numbers in arithmetic or mathematical problems grow larger. Also, maths anxiety can negatively affect working memory. It means that operations with extreme numbers can be limited for young children or students with maths anxiety. A good way to break the limitation is to find a simple way to operate with large numbers. We propose that using *powers of ten* as a notational tool may be the best way of achieving this goal.

Conventionally powers of ten are taught in the context of exponentials and logarithmic scales.  However, these abstractions are not necessary to grasp the basics. In the research described in this paper, we introduced *powers of ten*, using appropriate child-friendly language. We avoid the conventional mathematical language normally introduced in middle school. *Powers of ten* is introduced as a method for allowing simple calculations with big numbers. First, we explore the rules of powers of ten notation with examples like 10 x 10 = 100, where the rule is addition of the numbers of zeros. It easily follows that to multiply million ($10^6$) and billion ($10^9$) we just need add 6+9 for which we can choose whether to use to word quadrillion or simply use $10^{15}$. Simple large number arithmetic then becomes accessible for primary school students as we demonstrate in this paper.

**_Powers of ten for estimation._** From a mathematical point of view, powers of ten notation always involves approximation in the form of rounding to the chosen number of significant figures. It, therefore, is closely related to the concept of Fermi estimation. In general Fermi, estimation is associated with the phrase *it is better to be approximately right than exactly wrong* (Ratcliffe, 2007). It makes use of reasonable assumptions and evaluations to discover approximate answers. *Powers of ten* thinking at the single significant figure level is consistent with and facilitates Fermi estimation and allows students to estimate quantities that they never would have imagined to be within their ability.

Fermi estimation is about avoiding unnecessary information beyond our capability to comprehend. For example, the speed of light *c* is =299 792 458 m/s. In this rare case c is a *defined* quantity so it happens to be exactly right, but in general, nine significant figures are completely unnecessary for comprehension. In general, comprehension is facilitated by minimising the number of significant figures. For this reason, we choose c = 300000000 m/s, or $3 \times 10^8$ m/s as an appropriate representation of the magnitude of the speed of light for primary school students.

**_Potential errors of estimation_**. In adopting an approach to large numbers based on single significant figure estimations we must deal with one potential pitfall, which results from products of numbers. It is easy to see this with examples. When evaluating large numbers that are the products of several quantities, we must deal with cases such as a light year (the distance lights travel in one year) which in meters, by conventional arithmetic would be 299792458 (speed of light) x 31556952 (number of seconds in a year). By using powers of ten to 1 significant figure, the answer is 3 x $10^8$ (speed of light) x 3 x $10^7$ (seconds in a year) = 9 x $10^{15}$ meters. Now the student has a choice between 9 x $10^{15}$ or $10^{16}$. The user of Fermi estimation must make a judgement here. We often invoke a rule that 9 is approximately 10, but in some cases, estimations may become wrong by a whole order of magnitude due to propagation of errors.

At the simplest level it can be beneficial to work with zero significant figures, using sums like $10^3$ x $10^5$ =$10^8$, but at a more advanced level students need to explore the transition between sums like 3 x 3 is approximately $10^1$, 9 x 9 is approximately $10^2$. A good example is to calculate the number of seconds in a year: if you use 60 x 60 x 24 x 365 you get the number quoted above, but if you simplify it to $10^2$ x $10^2$ x 10 x $10^3$ you obtain $10^8$ which is an overestimate by a factor of 3.

In the next section we present the learning sequence that we used to develop logarithmic reasoning.

## The learning sequence.

The learning sequence consists of five key steps as shown in Figure 2. First, we give students an experience of exponential phenomena based on powers of two. We use powers of two to give students an intuitive understanding of exponential processes by discovering how doubling can quickly take us to extreme numbers. The next step is to connect powers of two to powers of ten. We use activities to show that ten powers of two is approximately equal to three powers of ten. This allows students to appreciate approximation tools and become familiar with the rules of powers of ten arithmetic, all through appealing activities. Finally, we introduce the entire scale of the Universe with powers of ten, by encompassing the entire universe within a 120 page book in which each page represents a power of ten.

**Fig.2**

*Five steps towards the understanding of the whole scale of the Universe with powers of ten.*

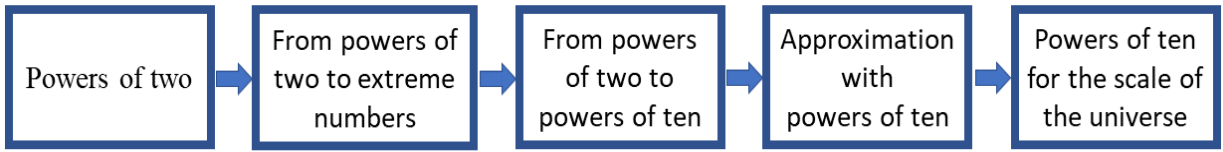

*1)* The 5 conceptual steps are taught through 13 activities listed in Table 2. **Powers of two (Activities 1-3).** Halving and doubling operations are natural for children because their

**Fig.3**

*Explore a) exponential growth with rice on a chessboard b) cutting measure tape in half 10 times to reach a piece nearly 1 mm long.*

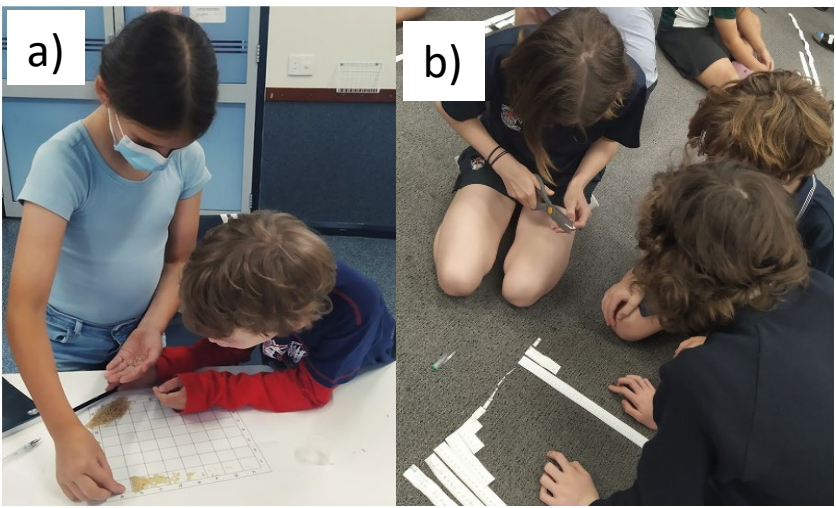

imagination is linked to experiences such as halving a piece of cake or doubling your pocket money. These operations are ideally suited for activity-based learning, while the activities can be described mathematically using powers of two (Fig.3). Halving and doubling allows easy introduction to the power function and the associated short-hand notation such as 2x2x2x2x2=$2^5$.

**2)** **From powers of two to extreme numbers (Activities 4-5 and 7 - 8).** To introduce numerically extreme numbers to students, we move step by step from numbers they are familiar with to extreme numbers using powers of two. This fosters a sense of proportionality. Doubling also allows students to conceptualise the rapid increase in numbers associated with pandemics, cell division, quantum computer power and population growth.

**3)** **From powers of two to powers of ten (Activities 6-8).** The concept of powers of two creates a simple bridge to understanding powers of ten through the approximation $2^{10} \sim 10^3$, and its reciprocal form $2^{-10} \sim 10^{-3}$. This approximation is realised by cutting 1m paper tape measures into halves ten times as shown in Figure 3. The tenth piece is very close to one millimeter. Students verify this by cutting a second tape measure into tenths. They can immediately observe the rough equality, as well as the exponential form represented by the paper histogram.

Students learn that repetitive multiplication or division ten times leads them to a rapid increase or decrease in number following the same procedure as doubling, but much quicker. This solid foundation is designed to facilitate future learning and comprehension of power functions with any base in the high school years *Western Australian Curriculum and Assessment Outline, 2023).* By asking students to label the paper pieces with both numbers and powers of ten notation, this is the first step towards proficiency in exponential notation.

**4)** **Approximation with powers of ten (Activities 9-10).** The fourth level of our program develops approximation skills. We encourage students to approximate to the closest power of ten. This allows them to discover how calculations with extreme numbers are simple with powers of ten arithmetic, such as multiplying one million ($10^6$) by one billion ($10^9$) by adding 6+9 to find the answer $10^{15}$. By analysing approximations with one significant figure, they develop an understanding of significant figures and approximation errors.

**5)** **Powers of ten for understanding scale of the universe (Activities 12 and 13).** The fifth step introduces the powers of ten (logarithmic) number line. Then they learn how to use powers of ten as a tool to place the enormous range of magnitudes, from atoms to the edge of the Universe on one number line. From a formal point of view, we are teaching logarithmic thinking.

The Powers of the Universe activity book is used to reinforce logarithmic thinking and understand extreme scales. We specify only SI units (meters, seconds, kilograms). Each page is a power of ten, numbered from -37 to 90. Students use search tools to find sizes of things they have heard about, such as the size of a quark (page -18) or atoms (page -10) up to the size of the observable Universe (page 26). People are on page zero, viruses on page -8. On other rows they add speeds: toe nail growth speed and continental drift speed are on pages -10 and -9 respectively, while the "speed limit of the universe" (light speed) is on page 8. Masses, time, frequencies, concentrations, and number counts can also be added to the book. The number of photons in the universe is on page 90, while the number of protons in the Sun is on page 57. Through the idea that mathematics has given them the power to encompass the entire universe in a small book, this activity helps students develop a logarithmic sense of scale.

**Table 2.**

*Five conceptual steps (each represented by a different colour) of learning progression taught with 13 activities*.

| | |
|---|---|
| **1. Ancestor Counts.** <br><br> *Doubling* | Using powers of two, students find the number of ancestors in any generation. Students create an imaginary family tree in classroom, asking their classmates play the role parents, grandparents etc. Starting with themselves and then doubling to represent their parents, the continue to double for each subsequent generation until they run out of classmates to represent ancestors. |
| **2. Knock-out (single-elimination) tournament.** <br><br> *Halving* | Powers of two are employed to determine the number of players in each round of a sports tournament. Students play rock-paper-scissors in pairs. Only winners progress to the next round, while the number of players halves each round.  Students play until one player is crowned the overall winner. |
| *3.* **Epidemics.** <br><br> ***Doubling and Exponential growth in time*** | The concept of exponential grow for epidemics is demonstrated with powers of two. The "virus" is transmissible to two people per exposure period. Before the activity begins, one student is marked as "sick" by the teacher. Students freely walk around the classroom and shake each other's hands. When shaking hands, the 'sick' student lightly scratches the other person's palm to "transmit" their virus.  Each repeated instance of group handshaking will be counted as the number of periods of exposure, for example 24 hours on the fourth period of exposure, all "sick" students will be asked to raise their hands. At this point, every student in the classroom will have the virus. Students then estimate how many days it would take to infect the entire population of real cities after infection begin. |
| **4. The mathematician tricks the emperor. Universe on the chessboard.** | Students investigate an exponential grow up to astronomical numbers, they are given a chessboard, a cup of rice and a spoon. Starting with one grain of rice in the first square, students double the number of grains in every square of the chessboard (Demi, 1997). They will most likely stop at square number 10, which |
| **Doubling, Exponential growth and extreme numbers** | requires 1024 grains of rice. Students write powers of two on every square. |
| **5. Tape measure halving activity: How small is an atom?**<br><br>*Halving and Exponential reduction to extreme small numbers* | Students investigate exponential decrease of size with powers of two by halving the one-meter tape measure ten times repeatedly. They count the number of pieces after every cut if they would cut every piece. Every level of cutting (Fig 2) represents the indices of powers of two, while the number of pieces is equal to the equivalent ordinary number. Students imagine continuing this experiment until the paper decreases to the size of an atom (~32 times). |
| **6. From powers of two to powers of ten**<br><br>*Transposing between powers of 2 and powers of ten* | Students investigate that $2^{10} \sim 10^3$. They cut a one-meter paper tape measure into repeated halves 10 times. Then they cut another one-meter paper tape measure into tenths 3 times to have a nearly same sized piece. Then this exercise is mentally extrapolated to find how many times this process needs to be continued to reach the size of the atom cutting into tenth comparing the process with cutting into halves. |
| **7. Doubling and populations**<br><br>*Comprehension of extreme numbers in the modern world* | We encourage students to use Google to determine the current global population and observe its ongoing growth trajectory. With knowledge about transposing between powers of 2 and powers of ten students return to ancestors' activity. For example, they compare the whole population of the world (about $10^8$) in the year 1000 with population of their own ancestors at this year. |
| **8. Rice on chessboard and epidemics in powers of ten**<br><br>*Powers of ten and doubling processes* | To reinforce *Transposing between powers of 2 and powers of ten* Students rethink activities 1 and 4. They estimate how many days it would take to infect the entire population (written in powers) of ten of their real cities.<br>They link certain squares on the chessboard to extreme known numbers in physics, such as the age of the universe in seconds or the distance to the closest star in metres using powers of ten and powers of two. |
| **9. How big is one million dollars? –**<br><br>*Comprehension of extreme numbers in the modern world* | This task challenges students to visualise the magnitude of an extremely large number using smaller numbers they are more familiar with. Students creating an imaginary shopping list, with the goal of successfully spending a large amount of money (for example, a million dollars) on realistically priced items.<br>To broaden the activity, students are encouraged to collect approximate population data for countries to understand how they contribute to whole world population. |

| 10.    Drama play: *Ten times Alice.* *Powers of ten proportional reasoning* | Students learn the "ten times" rule with simple poetry. In the drama, one student plays the role of Alice, the main character in *Alice in Wonderland.* After eating a magical biscuit her height becomes ten times bigger or ten times smaller.  Starting with their height they shrink to the size of an atom and grow in size up to the distance to the Moon. |
|---|---|
| 11.    Lazy maths: powers is faster. *Arithmetic with powers of ten* | Maths with powers is easy but not exact. Students practice "fast" calculations to find, for example, *how many seconds in a year* or *How much time is required to travel to the moon at the speed of a car?* They also have a competition in writing numbers with powers of ten and with zeros. |
| 12. Powers of the Universe book *Scale of the Universe with powers of ten* | Putting the Universe into a single book. A long-term activity consisting of the creation of a 130-page book called *Powers of the Universe.* All numbers in SI units describing the known Universe can be placed in this book, if each page is an increasing power of ten. |
| 13. Powers of ten dice *Scale of the Universe and arithmetic with powers of ten* | Powers of ten dice is a table game for familiarising students with the scale of the Universe using powers of ten. It combines knowledge of scales with practice in multiplying and dividing extreme numbers. |

Every step of our sequence provides skills and knowledge to develop further understanding easily. At the end of the program, students develop comprehension of the entire range of extreme numbers that describe our Universe supported by approximation skills and powers of ten thinking. We tested this novel learning sequence, using the methodology described below.

## Methodology

The methodology for the development and assessment of the *Powers of the Universe* module consisted of four main steps: a) creation of activities b) development sequence of lessons c) implementation of lessons, and d) assessment of students' outcomes.

**a)    Creation of activities.** The range of activities was designed to explicitly introduce, develop and reinforce mathematical concepts. The easily approachable activities are all designed to connect to real-life examples from science and everyday situations. The lesson structure prioritizes activities, around which the subsequent content is carefully developed.

b)    **Development of lessons sequence**. The learning strategy used in the program follows one that has been successfully developed by the Einstein-First project, which includes:

1. learning from group activities,

2. using models, toys, and games to illustrate concepts,

3. learning through role play and introducing historical content.

4. using Google search for data gathering where appropriate

A typical 1.5-hour lesson included three parts 1) Mathematical activities, experiments or games; 2) presentation of mathematical concepts related to physics learning or relevant to modern life; and 3) questions or quizzes for concept reinforcement.

**Implementation of lessons.** The 6-hour sequence of lessons was implemented at three of the project's partner schools (Table 3). We refer to the three trials as trial 1, trial 2 and trial 3 which correspond to the three schools.

**Table 3.**

*Lesson structure and duration of the trials in 1) and 2) government schools 3) private school with a stable multi-age class, taught different age groups*.

|   | Time | Duration | Duration of lesson | Number of lessons | Number of students | Students level |
|---|------|----------|--------------------|--------------------|--------------------|----------------|
| 1 | 6 hours | 1 month | 1.5 hours | 4 | 27 | Year 5-6 |
| 2 | 6 hours | 1 month | 1.5 hours | 4 | 19 | Year 5-6 |
| 3 | 6 hours 40 min | 2 months | 50 minutes | 8 | 20 | Year 3-7 |

c) **Assessment of students' outcomes.**

The assessment of the trials followed established procedures developed by the Einstein-First project (Authors 2023). This approach encompasses similar questionnaires, being administered both before and after the trial. Typically, the post-test included additional questions directly linked to the activities covered during the course. These questionnaires were developed by the authors, and reviewed by a panel of experts including experienced researchers and consultant school teachers. After initial trials some questions were revised. In the assessment process, students were allocated approximately 25 minutes to respond to eight questions in both the pre- and post-course questionnaires (Appendix). The knowledge questions include tasks for estimation, calculation, and open-ended questions as appropriate. The questionnaires aim to find out if there are natural difficulties in grasping certain concepts. Ten months after the trials we had the opportunity to re-test 22 of the participants to evaluate long-term retention and learning outcomes based on a simple test with five questions (See the Appendix).

We developed this module following the Model of Educational Reconstruction (MER), a well-established framework in science education that is used to examine the efficacy of teaching specific scientific concepts (Duit et al., 2012). Student outcomes from the first trial were monitored to allow assessment and improvement of the module when necessary. Additionally, to gather qualitative data regarding activities and lessons, interviews were conducted with both the teacher and the coordinating teacher. The primary enhancement of module in trials 2 and 3 involved revising the questionnaires, which encompassed the introduction of a visual family tree structure (Q1 Appendix) and instructions on employing powers of ten in question Q2 (Appendix). This process led to the addition of an activity entitled *"Lazy Math: Powers is faster"* in trials 2 and 3 (in the second activity time slot) to foster skills in performing operations involving powers of ten.

Because in general we were introducing entirely new material, we were interested in observing strong changes in students' knowledge and understanding, that we hoped would be observable in histograms without need for statistical analysis. Our sample sizes, age ranges and inhomogeneity of our samples also makes statistical inferences unreliable. For this reason, the data analysis that follows in the next section focuses on conceptual understanding and detailed analysis of questionnaire outcomes. Furthermore, as will be evident in the results section, the significance of the outcomes is readily apparent.

## Results and Discussion.

The questions in the tests were divided into six groups according to expected outcomes:1)comparison of measure units (metre and millimetre), 2) doubling and halving activity using powers of two, 3) writing numbers and multiplication with powers of ten, 4) scale of the Universe and Powers of ten book 5) overall scores and age-dependent factors for students 6) retention assesment.

**Comparison of measuring units.** Question 1 (Q1) was designed to assess students' knowledge of metric units using millimetres and meters, which we expected some students to have learnt in normal lessons. In the pretest, 55% of the students gave correct answers, increasing to 80% in the post test.

**Doubling and halving activity using powers of two**. The first two activities of the trials aimed to introduce students to the concepts of powers of two through doubling and halving. As shown in Fig 5a), before the three trials, only 26%, 53% and 5% students were able to calculate their number of fifth-generation ancestors(Q2). The scores rose to 96%, 95% and 85% respectively after the trials. The comparatively high pre-test score from trial 2 was due to the regular class

**Fig.5**

*Tests results for trials 1,2,3 a) doubling for counting ancestors(Q2) b) tape measure halving(Q3)*

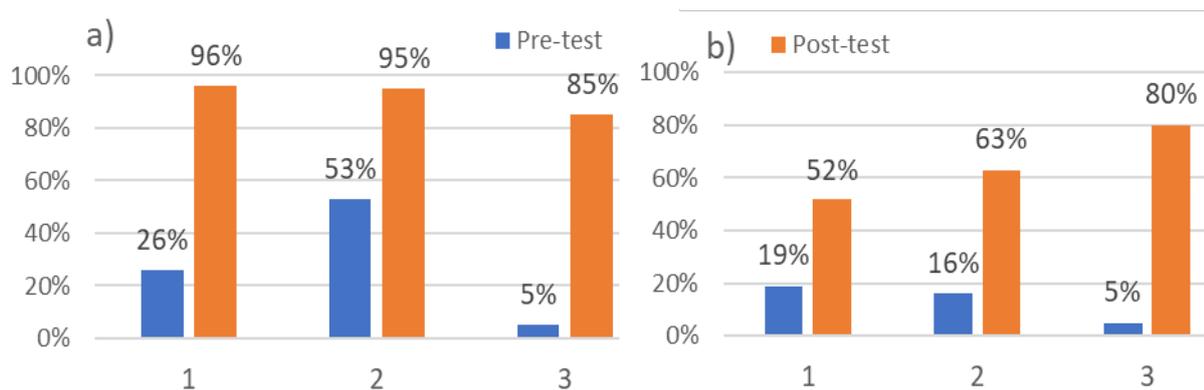

teacher having introduced doubling in an advanced mathematics lesson. The lower pre-test marks for trial 3 is thought to be due to the younger average age of this class. The average age for trial 3 was 9, whereas in trials 1 and 2, it was 11 years of age. Because a few students experienced difficulty connecting the number of generations to the words "great-great-great grandparents" we included a picture with the structure of a family tree for trials 2 and 3. However, the data indicated no significant change in scores.

Question Q3 asked how many halvings of a one-meter tape measure are needed to reach a size of roughly 1 mm. As shown in Fig.5b, the pre-test scores of 19 %, 16 % and 5% increased to 52%, 63% and 80% after the cutting activity. While all students had been actively engaged in the activity, it was clear from trial 1 that almost half the students had missed the key mathematical outcome: *10 cuttings in half takes us from 1m to approximately 1 mm*. In trial 2 the mathematical outcome was emphasised. Trial 3 extended this emphasis, by including estimations regarding rice on a chessboard, pandemics, and world population. A common mistake of students was related to their understanding of the idea of approximation. Some students halted their halving process after 7 halvings, deeming the size to be sufficiently small. After highlighting the importance of estimation accuracy and providing more context, post-test results in trial 2 and 3 improved to 63% and 80% respectively. The improvements from trials 1 to 3 indicates that with adequate reinforcement and context related activities, a majority of students were able to demonstrate competency in doubling, halving and powers of two, including the connection between powers of two and powers of ten after only about one hour of learning.

1. **Writing numbers and multiplication with powers of ten**. Question four (Q4) required students to express large numbers like ten thousand, million, and billion using either powers of ten or standard numerical notation. In trial 1 no guidance on employing powers of ten was given. The question assesses if students prefer using powers of ten as a shorthand way for writing numbers. Fig. 6a shows that only 11% used powers of ten in the pre-test, but after Trial 1, 78% adopted powers of ten for shorthand number notation. We intentionally instructed students to use both methods for number representation in Trials 2 and 3, resulting in 84% and 95% correct answers in the post-tests, compared to 68% and 35% in the pre-tests, respectively. The majority (68%) of Year 5 -6 students in trial 2 were already aware of the correct use of powers of ten from their regular classes. Their knowledge had been slightly improved after the trial.

Question five (Q5) was about the multiplication of powers of ten. The concept of multiplication was not explicitly taught in trial 1, but many students understood it through simple examples such as 10 x 10. In the post-test 44% of the students used addition of powers for multiplication of big numbers compared with only 4% in pre-test (Fig.6b). Clearly the majority were not

**Fig.6**

*Tests results for trials 1,2 and 3 a) using powers of ten for writing numbers b) application of powers of ten for multiplication.*

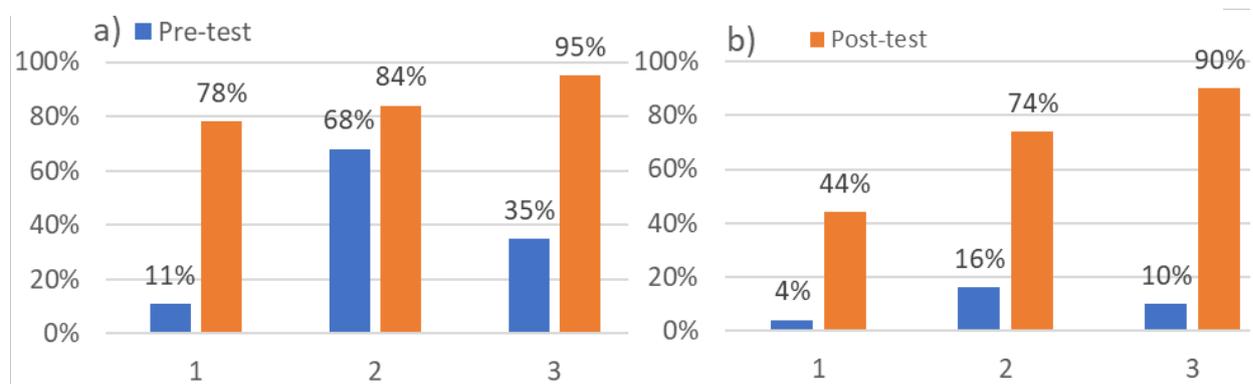

grasping multiplication by adding powers, so Activity 11, *Lazy Maths: Powers is faster* was added in trial 2 to reinforce this learning outcome. In trial 3, we extended this by including multiplication with powers from the *Powers of the Universe* book. Consequently, the post-test scores showed a significant improvement, reaching 74% and 90% in trials 2 and 3, compared to the initial pre-test scores of 16% and 10%, respectively (Fig.6b).

The above outcomes indicate that powers of ten arithmetic are accessible for children, but providing sufficient practice played a key role for establishing this learning outcome.

2. **Scale of the Universe.**

To assess students' knowledge of the scale of the Universe, we asked about the smallest and largest things students can think of (Q6). The students in three trials were categorized into two groups based on their pre-test responses. In the pretest, 60% of the students demonstrated a

**Fig. 7**

*a) Percentage of students defining extreme objects' range in broad and narrow scale. b) Using*

*Powers of the Universe book as a tool for understanding scale of the Universe after trials 1,2,3*

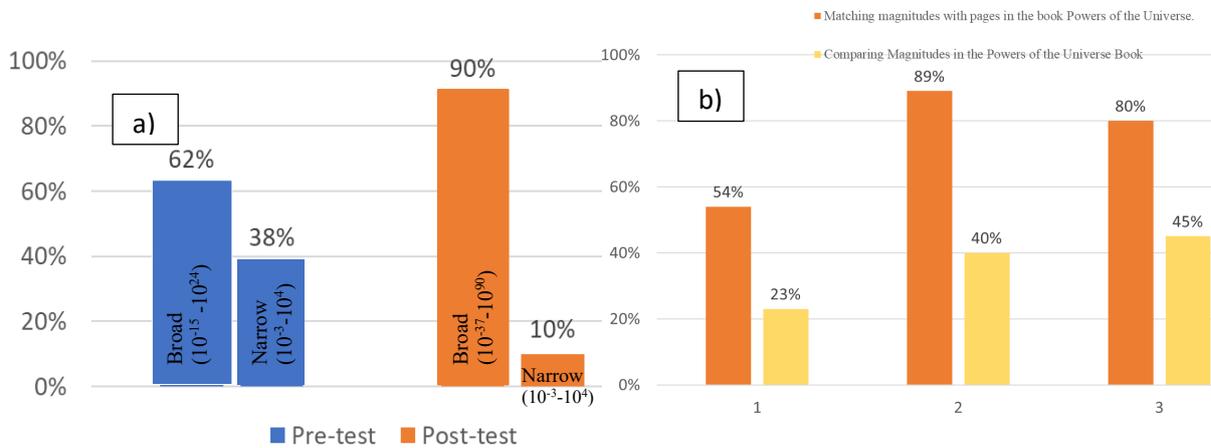

broad knowledge of scale, mentioning objects from both the atomic and cosmological scales (Fig

7a) with the most common responses being *atom* at 47% and *observable Universe* at 40%.

Surprisingly, 38% of students in the pre-test focused on a narrow scale (from $10^4$ to $10^{-3}$), with

answers like *ant, eraser*, *school*, reflecting objects closely connected to their personal

experiences. A sample of an answer from a student from trial 2:

*"The biggest objects: a house, a truck, a mountain.*

*The smallest objects: sesame seed, a grain of rice, a grain of sand."*

Susac (2014) describes this as *concrete mathematical reasoning* which excludes extremely large

or small objects because they remain abstract for them. After the trials the number of students

giving a narrow range had reduced to 10%: 90% gave answers encompassing atomic to

cosmological scales, with 50% mentioning subatomic particles. Overall, most of the students

indicated their ability to grasp a broader range of magnitudes. However, a few students (8%)

indicated number counts such as numbers of particles in the universe or in the Sun. The latter answers revealed the need for better precision in the question to clarify whether scale refers to linear dimension as opposed to other units.

To assess how students quantified extreme objects and magnitudes, in the post-test we presented a question Q7 related to Powers of the Universe book, to assess it's usefulness as a tool for understanding the scale, and for comparing magnitudes. According to the post-test, 54%, 89% and 80% of students, made a connection between the page of the Powers of the Universe book (powers of ten) and order of magnitude. In trial 1 only 23%, of students were able to compare objects by counting the number of pages between them(Q7b). Following trial 1 we extended the time for collaborative team work on the Powers of the Universe book by just 15 minutes. In trial 2 and trial 3, the scores increased to 40% and 45 %. Students' general enthusiasm for the Powers of the Universe book led us to believe that this is a highly beneficial activity but should be extended as a term-long activity.

5) **Overall scores and age-dependent factors for students.** The Figure 8 shows the percentage of correct answers in the Pre-test and Post-test, respectively, for every student in trial 3 starting with younger age. The results showed that some 7–9-year-olds with zero initial scores made a bigger improvement than students from the 10-13 years old group amongst this particular group of 20 students. In Figure 9 we represent the results of the pre-test and post-test for three trials. The histogram shows only a weak dependence of post-test scores on pre-test scores. The pre-test averaged at 18%, while the post-test reached 68%. Independently of age and knowledge before the trials, all the students had positive dynamics learning advanced mathematical concepts.

**Fig.9**

*The individual tests scores of students for three trials*

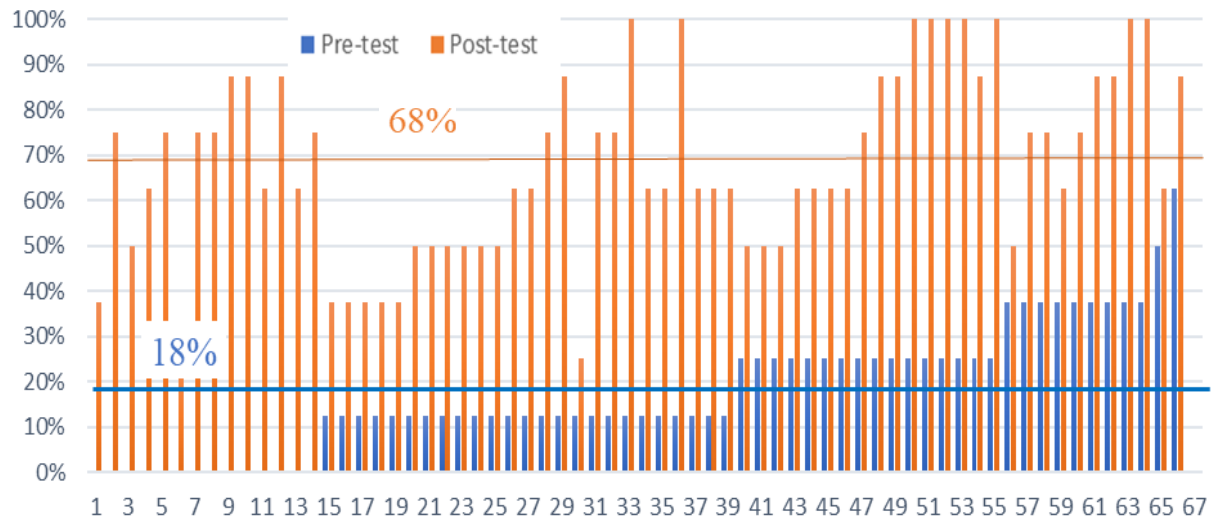

**Retention assessment.** We were able to re-test 22 students from trial 1 10 months after the trial.

Fig.10 a shows the answers to the question: What activities do you remember from the module ?

As can be seen from Fig.10a, the most popular activity that students remembered was the

Powers of the Universe book with 64% of students mentioning it. In second place was a role-play *Ten times Alice* with 41% of students remembering it. The third activity was the tape measure halving activity, which was mentioned by 32% of students. Large and small magnitudes that had been discussed with the Powers of the Universe book were mentioned by 27% of students, while about 18% of students remembered the ancestor counting activity. All students successfully listed one or more activities, offering detailed responses. Among these activities, the representation of the logarithmic scale with *Powers of the Universe* book emerged as the most memorable.

**Fig.10**

a) *Most remembered activities after 10 months: 1: Powers of the Universe book, 2: type measure halving activity 3: Ten times Alice 4: discussion about extreme objects 5: ancestor count b) the retention knowledge questions connected to these activities (Appendix).*

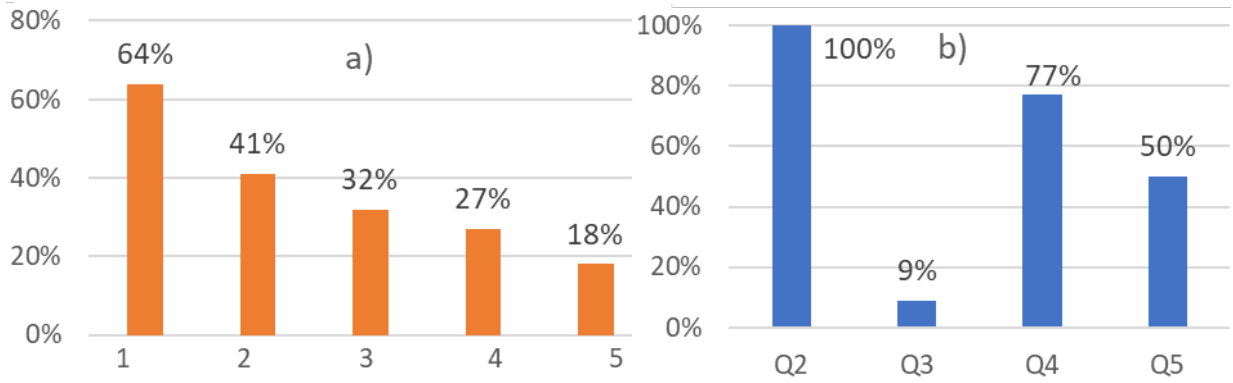

Fig.10b represents the retained knowledge of the concepts they learnt 10 months earlier. All students were able to write the numbers: billion, million, and ten thousand in powers of ten (Q2). Three quarters of students (77%) remembered the rules about dividing by ten as expressed in the songs in the *Ten Times Alice* play (Q4; Fig.10b). About one third of the class remembered the tape-measure halving activity but only 9%(Fig.10a) could correctly identify mathematical expression that $2^{10} \sim 10^3$ (Fig.10b). The most common mistake was *$10^3$ is equal to $3^{10}$* with 46% of answers. Half of the students (50%) could correctly connect magnitudes to page numbers in the Powers of Ten book. The common mistake in 23% of the answers was associating page number 0 with the number10.

If we make allowance for the above simple errors that could be easily remedied with more emphasis and more time, the score for Q3 increases to 55%, while Q5 reaches 73%. Overall, this result gives us confidence that small improvements and extended duration would give the majority of students in the age group tested a powerful grounding in logarithmic understanding of the universe.

Despite this positive conclusion we need to address the limitations of this study. Our results have been obtained from a small sample in schools that already had a positive attitude to improving students' science and mathematical ability. A larger trial across more diverse schools with more homogeneous age groups would be beneficial. As discussed above the results presented point to several improvements in emphasis that would improve future programs.

Our interventions are insufficient to draw conclusions regarding maths anxiety. Nevertheless, the observed engagement of students suggests that the approach being trialled here could be significant. In future trials following a more systematic implementation of our Maths for Einstein's Universe program, it would be very useful to implement attitude tests to determine the effectiveness of the program in reducing maths anxiety and encouraging more students to follow STEM career pathways.

This module was designed to be a first step towards a deeper mathematics education, by making students aware of the relevance and importance of mathematics prior to more formal and traditional studies. We suggest that a foundation comparable to the one presented here would yield favourable long-term results within the mathematical curriculum. The students participating in trial 3 are participating in a continuing *Powers of the Universe* book project planned to take two years. We will present results in upcoming papers.

## Conclusion

In this paper we have provided evidence that students between ages 7 and 13 can develop logarithmic thinking, understand exponential processes, position extreme numbers on a logarithmic number line, perform mathematical operations with them, and learn to represent the entire scale of the Universe through activity-based learning following the developed learning sequence.

Our findings from trials involving 66 students suggest that a strategically designed 6-hour sequence of activity-based lessons is an effective approach for learning extreme numbers and powers of ten. Approximately 90% of students were able to use powers of two to count ancestors and employ scientific notation to express numbers. In improved trials (trials 2 and 3), 80% of students confidently applied powers of ten for large number multiplication and effectively engaged with *Powers of the Universe, a book that represents the entire logarithmic scale of the universe*. A surprising result of our trials was that neither age (within the 7 to 13 range) nor prior knowledge played a significant role in achieving high levels of outcomes. We attribute this result to the accessibility of learning through activities. The fact that students retained a strong memory of the activities and knowledge acquired from them after 10 months indicates deep engagement and long-lasting impact from just 6 hours of learning.

## Appendix

**Pre- and Post-questionaries**

Q1. How many times bigger is 1 metre compared to 1 millimetre?

Q2. How many great-great-great-grandparents do you have?

Q3. How many times do you have to halve a 1 metre tape measure until you get to approximately 1 millimetre?

Q4. Write the following out either in numbers with zeros, or you may use powers of ten: Ten thousand\ One million\ One billion

Q5. What is the answer to ten to the power of 3 multiplied by ten to the power of eleven? $10^3 \times 10^{11} = ??$

Q6 Make a list of the three biggest/smallest things you can think of

Q7. In your Powers of the Universe book, the mass of Jupiter is on page 27 and the mass of the Sun is on page 30.

  a) How do you write the mass of the Sun in powers of ten

  b) Roughly how many Jupiters do we need to match the mass of the Sun? *Please give your answer using powers of ten.*

**Retention Questionnaire**

Q1. Last year you took a part in Einstein -First four-lesson program at your school. What activities do you remember?

Q2. There are two ways of writing big numbers. One is called powers of ten and the other just uses zeros. Please, write the following numbers in both ways.

  Ten thousand

  One million

  One billion

Q3. In one activity, a tape measure was divided and cut up in halves and then in tenths. From this cutting activity, you saw that three dividings of the tape measure by ten is nearly equal to ten dividings by _____________ . (fill the gap)

Q4. Please fill in the gaps

  Ten times smaller, take away a _______,

  Take away a ________,

  That's all you do,

  Ten times smaller, take away a________,

  One tenth Alice, that means you.

Q5. In your Powers of the Universe book you put in masses of the thing measured in kilograms. If you put in the mass of a blue whale on page 5, roughly how heavy is it? Please write it in powers of ten and with zeros.

## Funding

This research was supported by the Australian Research Council Linkage Grant LP 180100859 led by Professor David Blair at the University of Western Australia. The first author is a doctoral student who designed and implemented the *Maths for Einstein's Universe* program. The third author Professor David Treagust, from Curtin University is a member of the Einstein-First project.

## Authors contributions

The first author, Anastasia Lonshakova, is a doctoral student who designed and implemented the Maths for Einstein's Universe program, including the Powers of the Universe module. Professor David Blair is the project supervisor who provided key guidance for the design and implementation of the program, as well as the development of activities and questionnaires. Professor David Treagust, a specialist in educational research, notably contributed to the development of educational frameworks, questionnaires, and the effective presentation of ideas within the educational community. Professor Marjan Zadnik contributed to the effective presentation of results and ideas.

## Ethics Approval and Consent to Participate

The participants involved in this study gave their informed consent for this publication. The research was carried out under the University of Western Australia Ethics approval number 2019/RA/4/20/5875.

## Acknowledgements

The authors gratefully acknowledge the contribution of all the Einstein-First collaboration members:

David Wood for generously sharing his extensive experience in working with schools and educators, Dr Tejinder Kaur for coordinating the program and obtaining ethics permissions, Kyla Adams and Dr. Jesse Santoso for their advice during the development of ideas, Stefanie Palladino for developing the design of the Powers of the Universe book, Professor Ju Li for providing guidance and organizing support and others members of team Shon Boublil, Johanna Stalley, Dr. Elaine Horne for productive and enjoyable collaboration.

We would like to thank Peter Rossdeutscher and Professor Howard Golden, who have enabled us to raise donation funds to supplement our ARC Linkage funding (LP180100859) that allowed us to develop our on-line training programs. We also thank the ARC Centre of Excellence for Gravitational-Wave Discovery (OzGrav) for their continual support, especially in enabling us the develop school kits for our activities. We also thank our Einsteinian Physics Education Research (EPER) collaborators for their enthusiastic support. We are very grateful to the West Australian Department of Education for their support, the Independent Schools Association of Western Australia who have facilitated many of our trials, and the Science Teachers Association of Western Australia who have provided essential and continuous support. We are grateful to our participating schools' principals, teachers, and students for allowing us to conduct the program and for granting us permission to use their photographs and data for research purposes.

## References

Authors (2023)

Albarracín, L., & Gorgorió, N. (2019). Using large number estimation problems in primary education classrooms to introduce mathematical modelling. *International Journal of Innovation in Science and Mathematics Education*, *27*(2). https://doi.org/10.30722/IJISME.27.02.004

Albarracín, L., Ferrando, I., & Gorgorió, N. (2021). The role of context for characterising students' strategies when estimating large numbers of elements on a surface. *International Journal of Science and Mathematics Education*, *19*, 1209-1227

Ashcraft, M. H., & Krause, J. A. (2007) Working memory, math performance, and math anxiety. *Psychonomic Bulletin & Review*, *14*, 243-248.

Beswick, K. (2011). Putting context in context: An examination of the evidence for the benefits of 'contextualised' tasks. *International Journal of Science and Mathematics Education*, *9*, 367-390.

Boublil, S., Blair, D., & Treagust, D. F. (2023). Design and implementation of an Einsteinian energy learning module. *International Journal of Science and Mathematics Education*, 1-24. https://doi.org/10.1007/s10763-022-10348-5

Booth, J. L., & Siegler, R. S. (2006). Developmental and individual differences in pure numerical estimation. *Developmental Psychology*, *42*(1), 189.https://doi.org/10.1037/0012-1649.41.6.189

Bitterly, T. B., Van Epps, E. M., & Schweitzer, M. E. (2022). The predictive power of exponential numeracy. *Journal of Experimental Social Psychology*, *101*, 104347. https://doi.org/10.1016/j.jesp.2022.104347

Cheek, K. A. (2012). Students' understanding of large numbers as a key factor in their understanding of geologic time. *International Journal of Science and Mathematics Education*, *10*(5), 1047-1069.

Clary, R. (2009). How Old? Tested and trouble-free ways to convey geologic time. *Science Scope*, *33*(4), 62-66. https://doi.org/10.1126%2Fscience.1156540

Dehaene, S., Izard, V., Spelke, E., & Pica, P. (2008). Log or linear? Distinct intuitions of the number scale in Western and Amazonian indigene cultures. *Science*, *320*(5880), 1217-1220. https://doi.org/10.1126/science.1156540

Dehaene, S. (2003). The neural basis of the Weber–Fechner law: a logarithmic mental number line. *Trends in Cognitive Sciences*, *7*(4), 145-147. https://doi.org/10.1016/S1364-6613(03)00055-X

Duit, R., Gropengießer, H., Kattmann, U., Komorek, M., Parchmann, I. (2012). The Model of Educational Reconstruction – a framework for improving teaching and learning science. In: Jorde, D., and Dillon, J. (Eds) Science education research and practice in Europe. Cultural perspectives in science education, vol 5. Rotterdam: Sense Publishers. https://doi.org/10.1007/978-94-6091-900-8_2

Ellis, A. B., Ozgur, Z., Kulow, T., Dogan, M. F., & Amidon, J. (2016). An exponential growth learning trajectory: Students' emerging understanding of exponential growth through covariation. *Mathematical Thinking and Learning*, *18*(3), 151-181. https://doi.org/10.1080/10986065.2016.1183090

Harkness, S. S., & Brass, A. (2022). Large number placements on a bounded number line: The importance of explanations. *School Science and Mathematics*, *122*(2), 110-122.

Jones, M. G., Taylor, A., Minogue, J., Broadwell, B., Wiebe, E., & Carter, G. (2007). Understanding scale: Powers of ten. *Journal of Science Education and Technology*, *16*, 191-202.

Landy, D., Silbert, N., & Goldin, A. (2013). Estimating large numbers. Cognitive Science, 37(5), 775–799. https://doi.org/10.1111/cogs.12028.

Libarkin, J. C., Kurdziel, J. P., & Anderson, S. W. (2007). College student conceptions of geological time and the disconnect between ordering and scale. *Journal of Geoscience Education*,*55*(5), 413-422. https://doi.org/10.5408/1089-9995-55.5.413

Mahajan, S. (2018). The exponential benefits of logarithmic thinking. *American Journal of Physics*, *86*(11), 859-861. https://doi.org/10.1119/1.5058771

Maass, K., Geiger, V., Ariza, M. R., & Goos, M. (2019). The role of mathematics in interdisciplinary STEM education. *ZDM–The International Journal on Mathematics Education*, *51*, 869-884. https://doi.org/10.1007/s11858-019-01100-5

Moeller, K., Pixner, S., Kaufmann, L., & Nuerk, H. C. (2009). Children's early mental number line: Logarithmic or decomposed linear? *Journal of Experimental Child Psychology*, *103*(4), 503-515. https://doi.org/10.1016/j.jecp.2009.02.006

Nataraj, M. S., & Thomas, M. O. (2012). Student understanding of large numbers and powers: The effect of incorporating historical ideas. In J. Dindyal, L. P. Cheng, & S. F. Ng (Eds.), Mathematics education: Expanding horizons. Proceedings of the 35th annual conference of the Mathematics Education Research Group of Australasia (pp. 556–563). Singapore: MERGA.

Opfer, J. E., Thompson, C. A., & Furlong, E. E. (2010). Early development of spatial-numeric associations: evidence from spatial and quantitative performance of preschoolers. *Developmental Science*, *13*(5), 761-771. https://doi.org/10.1111/j.1467-7687.2009.00934.x

Ratcliffe, S. (Ed.). (2017). *Oxford essential quotations*. Oxford University Press.

Rajpaul, V. M., Lindstrøm, C., Engel, M. C., Brendehaug, M., & Allie, S. (2018). Cross-sectional study of students' knowledge of sizes and distances of astronomical objects. *Physical Review Physics Education Research*, *14*(2), 020108. https://doi.org/10.1103/PhysRevPhysEducRes.14.020108

Resnick, I., Davatzes, A., Newcombe, N. S., & Shipley, T. F. (2017). Using analogy to learn about phenomena at scales outside human perception. *Cognitive Research: Principles and Implications*, *2*(1), 1-17. https://doi.org/10.1186/s41235-017-0054-7

Resnick, I., Newcombe, N. S., & Shipley, T. F. (2017). Dealing with big numbers: Representation and understanding of magnitudes outside of human experience. *Cognitive Science*, *41*(4), 1020-1041. https://doi.org/10.1111/cogs.12388

Resnick, I., Davatzes, A., Newcombe, N. S., & Shipley, T. F. (2017). Using relational reasoning to learn about scientific phenomena at unfamiliar scales. *Educational Psychology Review*, *29*, 11-25. https://doi.org/10.1007/s10648-016-9371-5

Shepard, R. N., Kilpatric, D. W., & Cunningham, J. P. (1975). The internal representation of numbers. *Cognitive Psychology*, *7*(1), 82-138. https://doi.org/10.1016/0010-0285(75)90006-7

Siegler, R. S., Thompson, C. A., & Opfer, J. E. (2009). The logarithmic-to-linear shift: One learning sequence, many tasks, many time scales. *Mind, Brain, and Education*, *3*(3), 143-150. https://doi.org/10.1111/j.1751-228X.2009.01064.x

Stodolsky, S. S. (1985). Telling math: Origins of math aversion and anxiety. *Educational Psychologist*, *20*(3), 125-133. https://doi.org/10.1207/s15326985ep2003_2

Swarat, S., Light, G., Park, E. J., & Drane, D. (2011). A typology of undergraduate students' conceptions of size and scale: Identifying and characterizing conceptual variation. *Journal of Research in Science Teaching*, *48*(5), 512-533. https://doi.org/10.1002/tea.20403

Susac, A., Bubic, A., Vrbanc, A., & Planinic, M. (2014). Development of abstract mathematical reasoning: the case of algebra. *Frontiers in Human Neuroscience*, *8*, 679.

Trueborn, C. & Landy, D. Ordinal ranking as a method for assessing real-world proportional representations. *Proceeding 40th Annual Conference of the Cognitive Science Society* 2578–2583 (2018).

Thompson, C. A., & Opfer, J. E. (2010). How 15 hundred is like 15 cherries: Effect of progressive alignment on representational changes in numerical cognition. *Child Development*, *81*(6), 1768-1786. https://doi.org/10.1111/j.1467-8624.2010.01509.x

Tretter, T. R., Jones, M. G., Andre, T., Negishi, A., & Minogue, J. (2006). Conceptual boundaries and distances: Students' and experts' concepts of the scale of scientific phenomena. *Journal of Research in Science Teaching 43*(3), 282-319.
https://doi.org/10.1002/tea.20123

Ulusoy, F. (2019). Serious obstacles hindering middle school students' understanding of integer exponents. *International Journal of Research in Education and Science*, *5*(1), 52-69.

Vagliardo, J. (2006). Substantive knowledge and mindful use of logarithms: A conceptual analysis for mathematics educators. *Focus on Learning Problems in Mathematics*, *28*(3), 90.

Weber, K. (2002). Students' understanding of exponential and logarithmic functions. Second International Conference on the Teaching of Mathematics (pp. 1–10). Crete, Greece: University of Crete

Welch, A. G., & Areepattamannil, S. (Eds.). (2016). *Dispositions in teacher education: A global perspective*. Springer.

*Western Australian Curriculum and Assessment Outline (2023*) Retrieved October 5, 2023, from https://k10outline.scsa.wa.edu.au/